\documentclass[11pt]{article}
\usepackage{graphicx}

\usepackage{cite}
\usepackage{amsmath}

\usepackage{mathrsfs}
\DeclareMathAlphabet{\mathpzc}{OT1}{pzc}{m}{it}

\usepackage{geometry}
\begin{document}

\title{Impact of semi-annihilations on dark matter phenomenology --\\ an example of $Z_N$ symmetric scalar dark matter}

\author{G. B\'elanger$^1$, K. Kannike$^{2,3}$, A. Pukhov$^4$, M. Raidal$^3$}

\maketitle 
\noindent {$^1$ LAPTH, Univ. de Savoie, CNRS, B.P.110, F-74941 Annecy-le-Vieux Cedex, France \\
$^2$ Scuola Normale Superiore and INFN, Piazza dei Cavalieri 7, 56126 Pisa, Italy \\
$^3$ National Institute of Chemical Physics and Biophysics, R\"{a}vala 10,
Tallinn 10143, Estonia \\
$^4$  Skobeltsyn Inst. of Nuclear Physics, Moscow State Univ., Moscow 119992, Russia
}

\begin{abstract}
We study the impact of semi-annihilations $x_i x_j\leftrightarrow x_k X,$ where $x_i$ is any dark matter and $X$ is any standard model particle,  
on dark matter phenomenology.  We formulate minimal scalar dark matter models with an extra doublet and a complex singlet that predict non-trivial dark matter phenomenology with semi-annihilation processes for different discrete Abelian symmetries  $Z_N,$ $N>2$. 
We implement two such example models with $Z_3$ and $Z_4$ symmetry  in micrOMEGAs and work out  their phenomenology.  
We show that both semi-annihilations and annihilations involving only particles from two different dark matter sectors significantly modify the dark matter relic abundance in this type of models. We also study 
the possibility of dark matter direct detection in XENON100 in those models.
\end{abstract}

\section{INTRODUCTION} 

The origin of dark matter  of the Universe is not known. In popular models with new particles beyond the standard model particle content, such as the minimal supersymmetric standard model, an additional discrete $Z_2$ symmetry is introduced~\cite{Farrar:1978xj}. As a result, the lightest new $Z_2$-odd particle, $x,$ is stable and is a good candidate for dark matter. The phenomenology of this type of models has been  studied extensively.

The discrete symmetry that stabilises dark matter must be the discrete remnant of a broken gauge group \cite{Krauss:1988zc}, because global discrete symmetries are broken by gravity. The most natural way for the discrete symmetry to arise is from the breaking of a $U(1)_X$ embedded in a larger gauge group, e.g. $SO(10)$ \cite{Fritzsch:1974nn}. The latter contains gauged $B-L$ as a part of the symmetry, and the existence of dark matter can be related to the neutrino masses, leptogenesis and, in a broader context, to the existence of leptonic and baryonic  matter~\cite{Martin:1992mq,Kadastik:2009dj,Kadastik:2009cu}.  

Obviously, the discrete remnant of $U(1)_X$ need not to be $Z_2$ -- in general it can be any $Z_N$ Abelian symmetry. The possibility that dark matter may exist due to $Z_N,$ $N > 2,$ is a known \cite{Ibanez:1991hv,Agashe:2004ci,Agashe:2004bm,Dreiner:2005rd,Ma:2007gq,Agashe:2010gt,Agashe:2010tu,Batell:2010bp,DEramo:2010ep}, but much less studied scenario.\footnote{Phenomenology of $Z_3$-symmetric dark matter in supersymmetric models has been studied in Refs.~\cite{Ibanez:1991hv,Dreiner:2005rd} and in extra dimensional models in Refs.~\cite{Agashe:2004ci,Agashe:2004bm}.}
 Model independently, it has been pointed out in Ref.~\cite{DEramo:2010ep} 
  that  in $Z_N$ models the dark matter annihilation processes
 contain new topologies with different number of dark matter particles in the initial and final states -- called semi-annihilations --, for example $x x \leftrightarrow x^* X,$ where $X$ can be any standard model particle. It has been argued that those processes 
may significantly change the predictions for the dark matter relic abundance  in thermal freeze-out. 
Furthermore, an enlarged discrete symmetry group makes it possible to have more than one dark matter candidate.  In this case, annihilation processes involving only particles from the dark sectors, leading to  the
assisted freeze-out mechanism,
 can also influence the  relic abundance of both dark matter candidates ~\cite{Liu:2011aa,Belanger:2011ww}. The  assisted freeze-out mechanism in the case of a $Z_2\times Z_2$ symmetry was discussed in ~\cite{Belanger:2011ww}.
However, no detailed studies have been performed that compare dark matter phenomenology of different $Z_N$ models. 
This is difficult also because presently the publicly available tools for computing dark matter relic abundance do not  include  
the possibility of imposing a $Z_N$  discrete symmetry instead of a $Z_2$.

The aim of this work is to formulate the minimal scalar dark matter model that predicts different non-trivial scalar potentials for different $Z_N$ symmetries
and to  study their phenomenology. In particular we are interested in quantifying the possible effects of semi-annihilation processes 
$xx\leftrightarrow x^* X$ as well as of annihilation processes involving particles from two different dark sectors on generating the  dark matter relic abundance. In order to perform quantitatively precise analyses we implement 
minimal  $Z_{3}$ and $Z_{4}$ symmetric scalar dark matter models that contain one singlet and one extra doublet in micrOMEGAs~\cite{Belanger:2004yn,Belanger:2006is}.
 Using this tool we show that, indeed, the semi-annihilations and the annihilations between two dark sectors affect the dark matter phenomenology  and should be taken into account in a quantitatively precise way
 in studies of any particular model.

\section{$Z_{N}$ LAGRANGIANS}

\subsection{$Z_{N}$ symmetry}

Under an Abelian $Z_{N}$ symmetry, where $N$ is a positive integer, addition of charges is modulo $N$. Thus the possible values of $Z_{N}$ charges can be taken to be $0, 1, \ldots, N-1$ without loss of generality. A field $\phi$ with $Z_N$ charge $X$ transforms under a $Z_N$ transformation as $\phi \to \omega^X \phi$, where $\omega^N = 1$, that is $\omega = \exp (i 2 \pi/N)$.

A $Z_{N}$ symmetry can arise as a discrete gauge symmetry from breaking a $U(1)_{X}$ gauge group with a scalar, whose $X$-charge is $N$ \cite{Krauss:1988zc,Martin:1992mq}. For larger values of $N$, the conditions the $Z_{N}$ symmetry imposes on the Lagrangian approximate the original $U(1)$ symmetry for two reasons. First, assuming renormalisability, the number of possible Lagrangian terms is limited and will be exhausted for some small finite $N$, though they may come up in different combinations for different values of $N$. Second, if the $Z_{N}$ symmetry arises from some $U(1)_{X}$, the $X$-charges of particles cannot be arbitrarily large, because that would make the model nonperturbative. If $N$ is larger than the largest charge in the model, the restrictions on the Lagrangian are the same as in the unbroken $U(1)$.

We shall see below that in spite of the large number of possible assignments of $Z_N$ charges to the fields, the number of possible distinct potentials is much smaller.

\subsection{Field content of the minimal model}

In order to study the impact of different discrete $Z_{N}$ symmetries on dark matter phenomenology, the example model must contain more than one neutral particle in the dark sector.  The minimal dark matter model with such properties  contains, in addition to the standard model fermions and the standard model Higgs boson $H_{1}$, one extra scalar doublet $H_{2}$  and one extra complex scalar singlet $S$~\cite{Kadastik:2009dj}. 
In the case of $Z_2$ symmetry, as proposed in~\cite{Kadastik:2009dj}, those new fields can be identified with the well known
inert doublet $H_{2}$ \cite{Deshpande:1977rw,Ma:2006km,Barbieri:2006dq,LopezHonorez:2006gr} and the complex singlet $S$ \cite{McDonald:1993ex,Barger:2007im,Barger:2008jx,Burgess:2000yq,Gonderinger:2009jp}.
The phenomenology of those models is well studied. However, when both the doublet and singlet are taken into account,  qualitatively new features concerning dark matter phenomenology, electroweak symmetry breaking and collider phenomenology occur~\cite{Kadastik:2009cu,Kadastik:2009dj,Kadastik:2009ca,Kadastik:2009gx,Huitu:2010uc}.
The field content of the minimal scalar $Z_{N}$ model is summarised in Table~\ref{tab:fields}.

\begin{table}[htdp]	
\caption{Scalar field content of the low energy theory with the components of the standard model Higgs $H_{1}$ in the Feynman gauge. The value of the Higgs VEV is $v = 246$~GeV. }
\begin{center}
\begin{tabular}{cccccc}
Field & $SU(3)$ & $SU(2)_{L}$ & $T^{3}$ & $Y/2$ & $Q = T^{3} + Y/2$
\\
\hline
$H_{1} = \begin{pmatrix} G^{+} \\ \frac{v+h+iG^{0}}{\sqrt{2}} \end{pmatrix}$ & $\bf 1$ & $\bf 2$ & $\begin{pmatrix} \phantom{-}\frac{1}{2} \\ -\frac{1}{2} \end{pmatrix}$ & $\frac{1}{2}$ & $\begin{pmatrix} 1 \\ 0 \end{pmatrix}$ 
\\[3ex]
$H_{2} = \begin{pmatrix}
    -i H^{+}\\ \frac{H^{0} + i A^{0}}{\sqrt{2}}
    \end{pmatrix}$ & $\bf 1$ & $\bf 2$ & $\begin{pmatrix} \phantom{-}\frac{1}{2} \\ -\frac{1}{2} \end{pmatrix}$ & $\frac{1}{2}$ & $\begin{pmatrix} 1 \\ 0 \end{pmatrix}$  
\\[1ex]
$S = \frac{S_{H} + i S_{A}}{\sqrt{2}}$ & $\bf 1$ & $\bf 1$ & $0$ & $0$ & $0$ 
\\[1ex]
\end{tabular}
\end{center}
\label{tab:fields}
\end{table}

\subsection{Constraints on charge assignments}

The assignments of $Z_{N}$ charges have to satisfy
\begin{equation}
\begin{split}
  X_{S} &> 0, \\
  X_{1} &\neq X_{2}, \\
  -X_{\ell} + X_{1} + X_{e} &= 0 \mod N, \\
  -X_{q} + X_{1} + X_{d} &= 0 \mod N, \\
  -X_{q} - X_{1} + X_{u} &= 0 \mod N.
\end{split}
\label{eq:constraints:L}
\end{equation}
The first and second conditions arise from avoiding the $|H_{1}|^{2} S$ term and Yukawa terms for $H_{2}$, respectively, and the rest from requiring Yukawa interactions between $H_{1}$ and standard model fermions. The choice of $Z_{N}$ charges for standard model fermions, the standard model Higgs $H_{1}$, the inert doublet $H_{2}$ and the complex singlet $S$ must be such that there are no Yukawa terms for $H_{2}$ and no mixing between $H_{1}$ and $H_{2}$: only annihilation and
semi-annihilation terms for $H_{2}$ and $S$ are allowed. While we will see below that there are many assignments that satisfy Eq.~(\ref{eq:constraints:L}), in each case it was possible to find an assignment with the charges of standard model fields set to zero: $X_{q,\ell,u,d,e,1} = 0$.

All possible scalar potentials contain a common piece because the terms where each field is in pair with its Hermitian conjugate are allowed under any $Z_{N}$ and charge assignment. We denote it by $V_{\text{c}}$ (the `c' stands for `common'):
\begin{equation}
\begin{split}
V_{\text{c}}&= \lambda_{1} \left( |H_{1}|^2-\frac{v^2}{2} \right)^{2} + \mu_{2}^{2} |H_{2}|^{2} + \lambda_{2} |H_{2}|^{4} + \mu_{S}^{2} |S|^{2} + \lambda_{S} |S|^{4} \\
&+ \lambda_{S1} |S|^{2} |H_{1}|^{2}
+ \lambda_{S2} |S|^{2} |H_{2}|^{2} + \lambda_{3} |H_{1}|^{2} |H_{2}|^{2} + \lambda_{4} (H_{1}^{\dagger} H_{2}) (H_{2}^{\dagger} H_{1}).
\end{split}
\label{eq:V:c}
\end{equation}

\subsection{The $Z_{2}$ scalar potential}

There are 256 ways to assign the possible $Z_{2}$ charges $0,1$ to the standard model and dark sector fields. Of these, 8 satisfy Eq.~(\ref{eq:constraints:L}); among them, there are 2 different assignments to the dark sector fields: $X_{S} = X_{1} = 1, X_{2} = 0$ and $X_{1} = 0, X_{2} = X_{S} = 1$. Both give rise to the unique scalar potential
\begin{equation}
\begin{split}
    V &= V_{\rm c} + \frac{\mu_{S}^{\prime 2}}{2} ( S^{2}
    + S^{\dagger 2} ) + \frac{\lambda_{5}}{2} \left[(H_{1}^{\dagger} H_{2})^{2} + (H_{2}^{\dagger} H_{1})^{2} \right]\\
        &+ \frac{\mu_{S H}}{2} (S^{\dagger} H_{1}^{\dagger} H_{2} + S H_{2}^{\dagger} H_{1})
    + \frac{\mu'_{S H}}{2} (S H_{1}^{\dagger} H_{2} + S^{\dagger} H_{2}^{\dagger} H_{1})
    \\
    &+ \frac{ \lambda'_{S}}{2} (S^{4} + S^{\dagger 4})
    + \frac{ \lambda''_{S} }{2} |S|^{2} (S^{2} + S^{\dagger 2})
    \\
    &+ \frac{ \lambda'_{S1} }{2}
    |H_{1}|^{2} (S^{2} + S^{\dagger 2} )
    + \frac{ \lambda'_{S2} }{2}
    |H_{2}|^{2} (S^{2} + S^{\dagger 2} ).
\end{split}
\label{eq:V:Z2}
\end{equation}

\subsection{$Z_{3}$ scalar potentials and particle content}

There are 6561 ways to assign $0,1,2$ to the fields. Of these, 108 satisfy Eq.~(\ref{eq:constraints:L}); among them, there are 12 different 
assignments to the dark sector fields, giving rise to 2 different scalar potentials. The example potential we choose to work with (given by e.g. $X_{1} = 0, X_{2} = X_{S} = 1$) is
\begin{equation}
\begin{split}
V_{Z_3} &= V_{\text{c}} + \frac{\mu''_{S}}{2} (S^{3} + S^{\dagger 3}) + \frac{\lambda_{S12}}{2} (S^{2} H_{1}^{\dagger} H_{2} + S^{\dagger 2} H_{2}^{\dagger} H_{1}) \\
&+ \frac{\mu_{SH}}{2} (S H_{2}^{\dagger} H_{1} + S^{\dagger} H_{1}^{\dagger} H_{2}),
\end{split}
\label{eq:Lag:Z:3}
\end{equation}
which induces the semi-annihilation processes we are interested in.
The second one is obtained from Eq.~(\ref{eq:Lag:Z:3}) by changing $S \to S^\dagger$ (with $\mu_{SH} \to \mu'_{SH}$ and $\lambda_{S12} \to \lambda_{S21}$).

The following conditions are sufficient  to have the global minimum of potential
at electroweak vacuum  with $\langle S\rangle = 0, \langle H_2 \rangle =0$:
\begin{eqnarray}
  \lambda_1, \lambda_2, \lambda_S, \lambda_{S1}, \lambda_{S2} &>& 0\;, \\
  \lambda_3 + \lambda_4 &>&0 \;,   \\
  4 \lambda_{S1} \lambda_{S2} &>& \lambda_{S12}^2 \;, \\
   \frac{\mu^{\prime\prime 2}}{\lambda_S} + \frac{\mu_{SH}^2}{\lambda_3 + \lambda_4} &<& 4 \mu_S^2 \;. 
  \label{eq:vac:z3}
\end{eqnarray}
We use these conditions for our benchmark points. 

The last term in Eq.~(\ref{eq:Lag:Z:3}) induces a mixing between the  down component of $H_2$ and $S$.
In terms of the mass eigenstates  $x_1$, $x_2$, we have 
\begin{equation}
H_2=\left(  \begin{array}{c} -i H^+ \\ x_1 \sin{\theta} + x_2 \cos{\theta}\\
\end{array} \right),\;\;\;\;\; S= x_1 \cos{\theta}  - x_2 \sin{\theta}.
\end{equation}   
The dark sector of this model consists of 3 complex particles
$x_1$, $x_2$, and $H^+$ with the $Z_3$ charge  of 1.  Taking the  masses of $x_1$, $x_2$
and the mixing angle $\theta$ as free parameters of the model, we get the following relations 
\begin{eqnarray}
\mu_S^2   &=& M_{x_2}^2 \sin^2{\theta}+M_{x_1}^2 \cos^2{\theta}-\lambda_{S1} \frac{v^{2}}{2},\\
\mu_{SH}  &=&  -4 (M_{x_2}^2-M_{x_1}^2) \frac{\cos{\theta} \sin{\theta}}{\sqrt{2} v },\\
\mu_2^2   &=& -(\lambda_4+\lambda_3) \frac{v^2}{2} + M_{x_1}^2 \sin^2{\theta} +M_{x_2}^2\cos^2{\theta}.
\end{eqnarray}
The $\lambda_1$ and the  mass of  $H^+$  can be presented by formulas 
\begin{eqnarray}
\lambda_1 &=& \frac{1}{2} \frac{M_{h}^{2}}{v^{2}},\\ 
M_{H^+}   &=&  \sqrt{\mu_2^2+\lambda_3 \frac{v^2}{2}}.
\end{eqnarray}
where $M_h$ is mass of SM Higgs.
  
\subsection{$Z_{4}$ scalar potentials and particle content}

There are 65536 ways to assign $0,1,2,3$ to the fields. Of these, 576 satisfy Eq.~\eqref{eq:constraints:L}; among them, there are 36 different assignments to the dark sector fields, giving rise to 5 different scalar potentials.
Among those the only potential that contains semi-annihilation terms is
\begin{equation}
\begin{split}
V_{Z_4}^1 &= V_{\text{c}} + \frac{\lambda'_{S}}{2} (S^{4} + S^{\dagger 4}) + 
\frac{\lambda_{5}}{2} \left[(H_{1}^{\dagger} H_{2})^{2} + (H_{2}^{\dagger} H_{1})^{2} \right] \\
&+ \frac{\lambda_{S12}}{2} (S^{2} H_{1}^{\dagger} H_{2} + S^{\dagger 2} H_{2}^{\dagger} H_{1}) + \frac{\lambda_{S21}}{2} (S^{2} H_{2}^{\dagger} H_{1} + S^{\dagger 2} H_{1}^{\dagger} H_{2}),
\end{split}
\label{eq:Lag:3:Z:4}
\end{equation}
invariant under e.g. the assignment of $Z_4$ charges $X_{1} = 0, X_{2} = 2, X_{S} = 1$.

The following conditions are sufficient  to have global minimum of potential
at electroweak vacuum  with $\langle S \rangle = 0, \langle H_2 \rangle = 0$:
\begin{eqnarray} 
   \lambda_1, \lambda_2, \lambda_{S1},\lambda_{S2} &>& 0\,,\\
   \lambda_S - |\lambda'_S| & \geq& 0 \;, \\
   \lambda_{3} +\lambda_{4} - |\lambda_5| & > & 0 \;, \\
   (|\lambda_{S12}| + |\lambda_{S21}|)^2  &< & \lambda_{S1} \lambda_{S2}   . 
  \label{eq:vac:z4:2}
\end{eqnarray}
Our benchmark points considered below satisfy these conditions.

The other four scalar potentials can formally be obtained from the $Z_2$-invariant potential Eq.~(\ref{eq:V:Z2}) by setting all the new terms added to $V_{\rm c}$ to zero, with the exception of the
1) $\lambda'_S$, $\mu_{SH}$,
2) $\lambda'_S$, $\mu'_{SH}$,
3) $\mu'_S$, $\lambda'_S$, $\lambda''_S$, $\lambda'_{S1}$, $\lambda'_{S2}$,
4) $\mu'_S$, $\lambda'_S$, $\lambda''_S$, $\lambda'_{S1}$, $\lambda'_{S2}$, 
  $\mu_{SH}$, $\mu'_{SH}$ terms.

The $\lambda_5$ term in potential (\ref{eq:Lag:3:Z:4}) splits the down
component of $H_2$ into two  real scalar fields with different masses,
\begin{equation}
 H_2=\left(  \begin{array}{c} -i H^+ \\ \frac{H^0 + i A^0}{\sqrt{2}}\\ \end{array} \right).
\end{equation}
Note that the complex  scalar $S$ does not  mix   with $H_2$ because these fields have
different $Z_N$ charges. As a result this model contains two dark sectors, the first one with the complex scalar $S$ (the $Z_4$ charge is 1), 
the second one comprising the complex scalar $H^+$ and the real scalars $H^0$ and $A^0$ ( the $Z_4$ charge is 2). Any of the neutral particles with a non-zero $Z_4$ charge can be a dark matter candidate. 
We will consider the masses of the neutral  scalar particles, $M_S$, $M_{H^0}$ and $M_{A^0}$,  as independent
parameters, then  
\begin{eqnarray}
\mu_S^2 &=& M_S^2 - \lambda_{S1} \frac{v^2}{2},\\
\lambda_5    &=& \frac{M_{H^0}^2-M_{A^0}^2}{v^2}, \label{eq:lam5:z4}\\
\mu_2^2   &=& M_{H^0}^2- (\lambda_3+\lambda_4+\lambda_5) \frac{v^2}{2}, \\
M_{H^+}    &=& \sqrt{ \frac{M_{A^0}^2 + M_{H^0}^2 }{2}  - \lambda_4 \frac{v^2}{2}},\\
\lambda_1 &=& \frac{1}{2} \frac{M_{h}^{2}}{v^{2}}.
\end{eqnarray}

\section{RELIC DENSITY IN CASE OF THE $Z_3$ SYMMETRY}

\subsection{Evolution equations}

Consider the  $Z_3$-symmetric theory.
The imposed $Z_3$ symmetry implies, as usual, just one dark matter candidate.
This is because the $Z_3$ charges $1$ and $-1$ correspond to a particle and its anti-particle. 
The new feature is that processes of the type $xx\rightarrow x^* X$, where $X$ is any standard model particle, also contribute to dark matter annihilation. 
The equation for the number density reads
\begin{equation}
\frac{dn}{dt}=-\langle v\sigma^{x x^* \rightarrow  XX} \rangle  \left(n^2-\overline{n}^2 \right) -\frac{1}{2} \langle v\sigma^{xx\rightarrow
x^* X} \rangle
\left(n^2-n \, \overline{n} \right) -3H n,
\end{equation}
where we use $\overline{n}=n_{\rm eq}$, $H$ is the Hubble rate, and angular brackets mean thermal averaging. We define 
\begin{equation}
\sigma_v \equiv  \langle v\sigma^{x x^* \rightarrow XX} \rangle + \frac{1}{2} \langle v \sigma^{xx\rightarrow
x^* X} \rangle \quad {\rm and} \quad
\alpha=\frac{1}{2} \frac{\sigma_v^{x x\rightarrow x^* X}}{\sigma_v},
\end{equation}
which means that $0\leq \alpha\leq 1$. Here and in the following we use the notation,
$\sigma_v^{xx\rightarrow
x^* X}\equiv \langle v \sigma^{xx\rightarrow
x^* X} \rangle$. In terms of the abundance, $Y=n/s,$ where $s$ is the entropy density, we obtain
\begin{equation}
\frac{dY}{dt}=-s \sigma_v \left(Y^2-\alpha Y \overline{Y} -(1-\alpha) \overline{Y}^2 \right) 
\end{equation}
or, using the entropy conservation condition $ds/dt = -3 H s$,
\begin{equation}
3H\frac{dY}{ds}=\sigma_v \left(Y^2-\alpha Y \overline{Y} -(1-\alpha) \overline{Y}^2 \right).
\end{equation}
where $\overline{Y}=Y_{\rm eq}$ is the equilibrium abundance. We use standard formulae for $H(T)$ and $s(T)$  \cite{Gondolo:1990dk} that allow to 
replace the entropy evolution with the temperature one.   To solve this equation we follow the usual procedure~\cite{Gondolo:1990dk,Belanger:2004yn}. Writing $Y=\overline{Y}+\Delta Y$ we find the starting point for the numerical solution of this equation with the Runge-Kutta method using
\begin{equation}
3H\frac{d\overline{Y}}{ds}=\sigma_v \overline{Y}\Delta Y \left( 2-\alpha \right),
\end{equation} 
where $\Delta Y\ll Y$. This
is similar to the standard case except that  $\Delta Y$ increases by a factor $1/(1-\alpha/2)$.
Furthermore, when solving numerically the evolution equation, the decoupling condition
$Y^2\gg \overline{Y}^2$ is modified to 
\begin{equation}
Y^2\gg \alpha Y \overline{Y} + (1-\alpha) \overline{Y}^2.
\end{equation}
This implies that the freeze-out starts at an earlier time and lasts until a later time as compared with the standard case.
This modified evolution equation is implemented in micrOMEGAs~\cite{Belanger:2006is,Belanger:2010gh}.
Although semi-annihilation processes can play a significant role in the computation 
of the relic density, the solution for the abundance depends only weakly on the parameter $\alpha$, typically   only by a few percent.
This means in particular that  the standard  freeze-out approximation works with a good precision.

\subsection{Numerical results with micrOMEGAs}

Using the scalar potential defined in Eq.~(\ref{eq:Lag:Z:3}) we have implemented in micrOMEGAs the scalar model with a $Z_3$ symmetry.  The scalar sector contains an additional scalar doublet and one complex singlet. The neutral component of the doublet mixes with the singlet, the lightest component  $x_1$ is therefore the dark matter candidate, while  the heavy component $x_2$ can decay into $x_1h$, where $h$ is the standard model-like Higgs boson. Because $h$ can decay into light  particles, $x_{2}$ is unstable even if the mass difference between $x_{1}$ and $x_{2}$ is small. Note that the doublet component of DM 
has a vector interaction with the $Z$. This interaction is determined by the $SU(2)\times U(1)$ gauge
group and leads to a large direct detection signal in conflict with exclusion limits, for example from XENON100 \cite{Aprile:2011hi}.
The only way to avoid this constraint is to consider a DM with a 
 very small doublet component, namely  we have to assume that the mixing angle
\begin{equation}
\theta \le 0.025.
\end{equation}
  In the limit of small mixing,   annihilation  processes  such as  $x_1 x_1^* \to  XX$  where X stands for $W,Z,h$, are  dominated   by  the  $\lambda_{S1}|S|^2|H_1|^2$ term.  The semi-annihilation process  $x_1 x_1\rightarrow x_1^* h$  is mainly
determined  by  a product of  $\mu''_S$ and  $\lambda_{S1}$ arising from the terms
           $\mu''_S (S^3 +S^{\dagger 3})/2$ and $ \lambda_{S1} |S|^2 |H_1|^2 $ in Eq.~\ref{eq:V:c} and Eq.~\ref{eq:Lag:Z:3}.
To illustrate a scenario where semi-annihilation channels contribute significantly and which predicts  reasonable values for the relic density and the direct detection rate, we choose a benchmark point with the following parameters 

\begin{table}[htd]
\begin{center}
\begin{tabular}{cc|cc|cc|cc}
$\lambda_2$     &    0.1 & $\lambda_S $    &    0.2    & $\lambda_{S12}$ &    0.1 & $M_{x_1}$ &  150~GeV  \\
$\lambda_3$     &    0.1 & $\lambda_{S1}$  &   0.05  & $M_h$           &  125~GeV & $M_{x_2}$ &  400~GeV    \\
$\lambda_4$     &    0.1 & $\lambda_{S2}$  &    0.1  &$\mu''_S$  &  80 ~GeV  &$\sin{\theta}$ & 0.025     \\   
\end{tabular}
\end{center}
\caption{Benchmark point for $Z_3$.}
\label{tab:bench:z3}
\end{table}

\begin{figure}
\begin{center}
\includegraphics[height=8cm, width=15cm,angle=0]{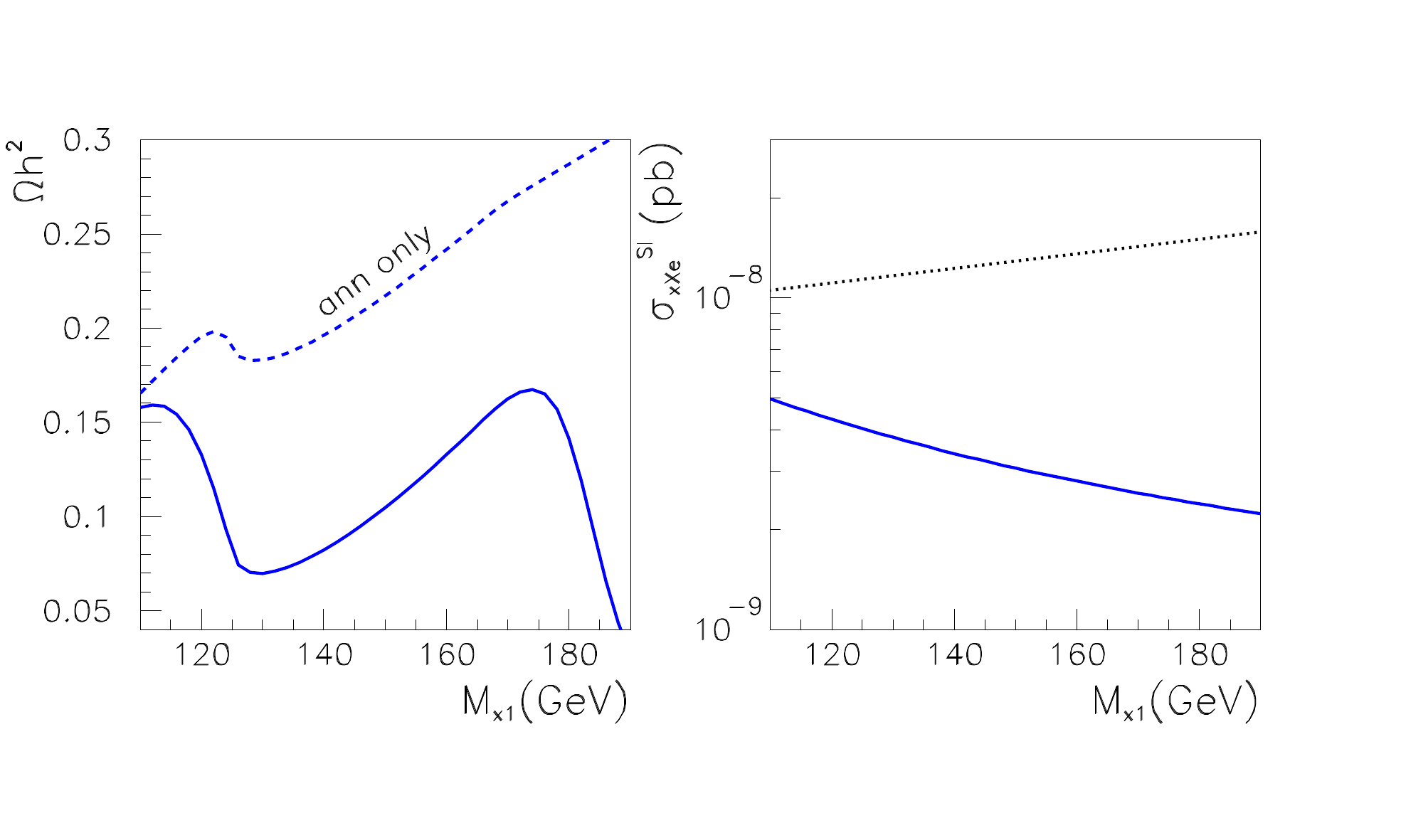}
 \vspace{-.8cm}
 \caption{(Left panel) $\Omega h^2$ as a function of the dark matter mass for the benchmark point with semi-annihilation (solid line), and   without semi-annihilation (dashed). (Right panel)  
$\sigma_{x_1 \mathrm{Xe}}^{\rm SI}$ (solid). 
The experimental limit from XENON100~\cite{Aprile:2011hi} is also displayed
(dashed). 
}
\label{fig:omega}
\end{center}
\end{figure}

For this point, the relic  density is $\Omega h^2 = 0.105$. The   dominant  contribution
to  $(\Omega h^2)^{-1}$ is from semi-annihilation (54\% for $x_1 x_1 \to h x_1^* $) while the annihilation channels 
 $x_1 x_1^*\to WW,ZZ,hh$ give a relative contribution of 22\%,13\% and 10\% respectively. 
Fig.~\ref{fig:omega} illustrates the dependence  of the relic density  on the DM mass  as compared to the relic density when semi-annihilation is ignored, $(\Omega h^2)_{\rm ann}$.   Here all other parameters are fixed to their 
benchmark values. When $M_{x_1}=110$~GeV, semi-annihilation with a Higgs in the final state is kinematically forbidden at low velocities. If $M_{x_1}$ increases, semi-annihilation plays an important role and  $\Omega h^2$ decreases rapidly due to the contribution of  the channel $x_1 x_1 \to h x_1^* $. Note that  $(\Omega h^2)_{\rm ann}$  also decreases when $M_{x_1}$ is such that the channel $x_1 x_1^* \rightarrow  hh$ is allowed.
When $M_{x_1}$ approaches  $M_{x_2}/2$, $\Omega h^2$ falls again because the semi-annihilation channel is enhanced due to  
$x_2$ exchange near  resonance.

The spin independent (SI) scattering cross section on nuclei as a function of the DM mass is illustrated in Fig.~\ref{fig:omega} (right panel). Here we average over dark matter and anti-dark matter cross sections assuming that they have the same density. The main contribution comes from the $Z$-exchange diagram because there is a $x_1 x_1^* Z$ coupling\footnote{In the inert doublet model with a $Z_2$ symmetry~\cite{Deshpande:1977rw,Barbieri:2006dq}, a $\lambda_5$ term splits the complex doublet into a scalar and a pseudoscalar, when the mass splitting is small such coupling leads to inelastic scattering.}.
Furthermore, one can easily show that the scattering amplitudes are not the same for protons and neutrons, with
$f_p=(4\sin^2\theta_W -1) f_n =- 0.075 f_n$. 
Since the current experimental bounds on $\sigma_{x p}^{\rm SI}$ are extracted from experimental results assuming 
that the couplings to protons ($f_p$) and neutrons ($f_n$) are equal and the
same as the couplings of $x_1^*$ to protons ($\bar{f}_p$) and neutrons ($\bar{f}_n$),  we define the normalised cross section on a point-like 
nucleus~\cite{Belanger:2010cd}:
\begin{eqnarray}\label{NormalizedCross}
\sigma_{x N}^{\rm SI} = 
\frac{2}{\pi} 
\left( \frac{M_N M_{x_1}}{M_N+M_{x_1}}\right)^2
  \left( \frac{[Zf_p + (A-Z)f_n]^2}{A^2}  + \frac{[Z\bar{f}_p + (A-Z)\bar{f}_n]^2}{A^2}  \right).
\end{eqnarray}
This quantity can directly be compared with the limit on $\sigma_{x p}^{\rm SI}$.

\section{RELIC DENSITY IN CASE OF THE $Z_4$ SYMMETRY}

\subsection{Evolution equations}

In the case of a  $Z_4$ symmetry  all particles can be divided into 3 classes\footnote{We take into account that $3 = -1 \mod 4$, so the particle with $X$-charge $3$ is the antiparticle of a particle with $X$-charge $1$.} 
\{0,1,2\}
according to the value of their $Z_4$ charges modulo $4$. We can choose SM particles to have $X_{\rm SM} =  0$. We will use the notation  $\sigma_v^{abcd}$ for the thermally
averaged  cross section for reactions $ab\to cd$  where $a,b,c,d=0,1,2$ represent any particle with given $X$-charge.   Let $M_{x_1}$ and
$M_{x_2}$  be the masses of the  lightest  particles of classes 1 and 2
respectively. The lightest particle of class 1 is always stable and therefore a DM candidate.
The
lightest particle of class 2 is stable and can be a second DM candidate if $M_{x_2} < 2 M_{x_1}$. 
Note that if $M_{x_2} > 2 M_{x_1}$, then $x_2$  will decay  before the freeze-out of $x_1$ and the relic density can be computed following the standard procedure.  

The equations for the number density of particles 1 and 2 read
\begin{eqnarray}
\frac{dn_1}{dt}&=&-\sigma_v^{1100}  \left(n_1^2-\bar{n}_1^2 \right) -
\sigma_v^{1120}\left( n_1^2- \bar{n}_1^2 \frac{n_2}{\bar{n}_2} \right)
- \sigma_v^{1122}\left( n_1^2- n_2^2 \frac{\bar{n}_1^2}{\bar{n}_2^2}
\right)- 3H n_1, \\
\frac{dn_2}{dt}&=&-\sigma_v^{2200}  \left(n_2^2-\bar{n}_2^2 \right)
+ \frac{1}{2}\sigma_v^{1120}\left(n_1^2- 
\bar{n}_1^2\frac{n_2}{\bar{n}_2} \right)
  - \frac{1}{2}\sigma_v^{1210}\left(n_1 n_2- n_1 \bar{n}_2
\right)\nonumber\\
&&-  \sigma_v^{2211}\left( n_2^2- {n}_1^2
\frac{\bar{n}_2^2}{\bar{n}_1^2} \right)  -3H n_2,
\end{eqnarray}
where we use  $\bar{n}_i$ to designate the equilibrium number  density of particle $x_i$. 
In $\sigma_v^{abcd}$ all annihilation and coannihilation processes are taken into account.
Here the semi-annihilation processes include all those, where 2 DM particles annihilate into one DM and one standard particle, specifically 
$\sigma_v^{1120}$ and $\sigma_v^{1210}$. These two cross sections  are also 
described by the same matrix element. However, there is no simple relation between these
two cross sections because one process is in the $s$-channel and the other in the $t$-channel.  In terms of the abundance, $Y_i=n_i/s$, 
\begin{eqnarray}
\label{z4eq1}
3H\frac{dY_1}{ds}&=&\sigma_v^{1100} \left(Y_1^2-\overline{Y}_1^2
\right) 
+   \sigma_v^{1120} \left(Y_1^2-Y_2 \frac{\overline{Y}_1^2}{\overline{Y}_2} 
\right) +\sigma_v^{1122}
\left( Y_1^2-Y_2^2 \frac{\overline{Y}_1^2}{\overline{Y}_2^2}\right),\\
\label{z4eq2}
3H\frac{dY_2}{ds}&=&\sigma_v^{2200} \left(Y_2^2-\overline{Y}_2^2
\right)
  -\frac{1}{2} \sigma_v^{1120} \left(Y_1^2-Y_2
   \frac{\overline{Y}_1^2}{\overline{Y}_2} 
\right) 
+ \frac{1}{2}  \sigma_v^{1210} Y_1 \left(Y_2-\overline{Y}_2 \right) \nonumber\\
&&+\sigma_v^{2211}
\left( Y_2^2-Y_1^2 \frac{\overline{Y}_2^2}{\overline{Y}_1^2}\right).
\end{eqnarray}
Solving these equations  we use standard  formulas for entropy $s(T)$ and the Hubble rate $H(T)$ temperature dependence \cite{Gondolo:1990dk}  that allow to replace the dependence on entropy with one on temperature.
The thermally averaged cross section involving particles of different sectors can be expressed as
\begin{eqnarray}
\sigma_v^{IJKL}(T) &=& \frac{T}{64\pi^5 s^2 \overline{Y}_I(T)\overline{Y}_J(T)}\int \frac{ds}{\sqrt{s}} K_1\left(\frac{\sqrt{s}}{T}\right) p_{\rm in}
p_{\rm out} \nonumber \\  &&  \sum_{\substack{a\in I\; b\in J\;\\ c\in K\; d\in L \\\mathrm{pol.}}}  \int^1_{-1} 
|\mathscr{M}_{ab\to cd}(\sqrt{s}, \cos{\Theta})|^2 d\cos{\theta},  \label{sigmaV} \\
\overline{Y}_I(T)&=&\frac{T}{2\pi^2 s} \sum_{ i\in I} g_i m^2_i K_2(\frac{m_i}{T}),
\end{eqnarray}
where $\mathscr{M}_{ab\to cd}$ is the  matrix element for the $2\rightarrow 2$ process and $K_1,K_2$ are modified Bessel functions of the second kind.  
 For reactions which are  kinematically open at zero relative velocity,   $\sigma_v$
 depends slowly  on temperature. Otherwise there is a strong 
$ \exp(- \Delta M/T)$ temperature dependence, where  $\Delta M$ is the 
difference between the sums of the masses of outgoing  and incoming  particles. 
Equation~(\ref{sigmaV}) leads to relations  between different cross sections 
\begin{equation}
       Y_I Y_J \sigma_v^{IJKL} = Y_K Y_L \sigma_v^{KLIJ}.
\end{equation}
In particular it implies that, $\sigma_v^{0211}=\sigma_v^{1120}Y_1^2/Y_2$, where the abundance of incoming SM particles 
$Y_0=1$.

Introducing $\Delta Y_i = Y_i - \overline{Y}_i$, Eqs.~(\ref{z4eq1})~and~(\ref{z4eq2}) take a simple form \begin{equation}
\label{MatrixZ4Eq}
        3H\frac{\Delta Y_i}{ds} = -C_i +A_{ij}(T)\Delta Y_j +Q_{ijk}(T) \Delta Y_j \Delta Y_k,
\end{equation}
where
\begin{eqnarray}
 C_i &=& 3H\frac{d \overline{Y}_i}{ds},\\
 A&=&\left( \begin{array}{cc}
2 (\sigma_v^{1100}+\sigma_v^{1122}+\sigma_v^{1120}) \overline{Y}_1 &-(\sigma_v^{1120}+2\sigma_v^{1122})\frac{\overline{Y}_1^2}{\overline{Y}_2}   \\
-\sigma_v^{1120} \overline{Y}_1 -2\sigma_v^{1122}\overline{Y}_1    &  2(\sigma_v^{2200}+\sigma_v^{2211})\overline{Y}_2
+ 0.5(\sigma_v^{1210} + \sigma_v^{1120}\frac{\overline{Y}_1}{\overline{Y}_2} )\overline{Y}_1
\end{array} \right), \\
 Q_1 &=& \left( \begin{array}{cc}
\sigma_v^{1100}+\sigma_v^{1122}+\sigma_v^{1120}  & 0 \\
0       & -\sigma_v^{2211}
\end{array} \right),\\
  Q_2&=&\left( \begin{array}{cc}
-\sigma_v^{1120} -\sigma_v^{1122} & \frac{1}{2}\sigma_v^{1210} \\
0 & \sigma_v^{2200}+\sigma_v^{2211} 
\end{array} \right). \\
\end{eqnarray}

At large temperatures we expect  the densities of  both DM components
to be close to their equilibrium values. In general in micrOMEGAs~\cite{Belanger:2001fz} the equation for the abundance is solved numerically starting from large temperatures. However, this procedure poses a problem for Eq.~(\ref{MatrixZ4Eq}). The step of the numerical solution is inversely proportional to $A(T)$  and as long as  $A(T)$ is not suppressed by the  Boltzmann factor included in $\overline{Y}$,   the step is too small and the numerical method fails. 

To avoid this problem,  we use the fact that at large temperatures 
one  can neglect the $Q$ term in Eq.~(\ref{MatrixZ4Eq}) and write the explicit solution for  the linearised equation.
The approximate solution in the case of large $A$ is 
\begin{equation}
    \label{largeTz4}
    \Delta Y_i(s)=  A^{-1}_{ij}(s) C_j(s).
\end{equation}
One can use Eq.~(\ref{largeTz4}) to find the lowest temperature where $\Delta Y_i \approx 0.05 Y_i $ and start solving numerically  
Eq.~(\ref{MatrixZ4Eq}) from this temperature. In the general case  it gives a   reasonable step for the numerical solution $\delta s/s \approx 0.1$, where $s$ is the variable of integration.  
This method can, however, lead to some numerical problems if  the masses of the two dark matter particles are very different. Let us call the light particle $\mathpzc{l}$ and the heavy particle $\mathpzc{h}$.
We have to start the  numerical solution at a  temperature $T$ above the  freeze-out temperature  of the heaviest DM, 
\begin{equation} 
T_{{\rm fo}\mathpzc{h}} \approx M_{\mathpzc{h}}/25.
\end{equation}
 At this temperature,
\begin{equation} 
\frac{Y_{\mathpzc{l}}}{Y_{\mathpzc{h}}} \approx \exp{\frac{M_{\mathpzc{h}}-M_{\mathpzc{l}}}{T_{{\rm fo}\mathpzc{h}}}},
\end{equation}
and the step in the numerical   solution of the  two component equations will be
suppressed by a factor $\exp{(-M_{\mathpzc{h}}-M_{\mathpzc{l}}) / T_{{\rm fo}\mathpzc{h}}}$. This small step size is problematic when solving numerically the equation with the Runge-Kutta method. This occurs when $M_{\mathpzc{h}}/M_{\mathpzc{l}} >2$. In this case
the equation for the heavy component must be solved independently
assuming that the light component has reached its equilibrium density.  
If $M_{\mathpzc{h}}/M_{\mathpzc{l}} <2$, the
Runge-Kutta procedure can be used to successfully solve the  thermal evolution equations~(\ref{MatrixZ4Eq}).

The abundances $Y_1$ and $Y_2$ will be modified by the interactions between the two dark matter sectors.\footnote{Note  that $Y_1$ and $Y_2$ correspond to the abundances of the particles with a given $Z_4$ charge. The relative size of the masses
of the DM particles depend on the choice of parameters in a given model.} 
Thus the new terms in Eq.~(\ref{z4eq1}) will simply add to the standard annihilation process with SM particles and will contribute to decrease the final abundance $Y_1$.  After $x_2$ freezes-out, interactions of the type $22\rightarrow 11$ lead to an increase of $Y_2$.
When $M_{x_1} \ll M_{x_2}$, the evolution of $Y_2$ will be strongly influenced by the first sector since at its freeze-out temperature $Y_1$ is large. Following the same  argument as above the new annihilation terms in Eq.~(\ref{z4eq2}) will  contribute to a decrease in the final abundance $Y_2$.
Furthermore, the semi-annihilation process $12\rightarrow 10$ which is always kinematically open means
that $x_1$ acts as a catalyst for the transformation of $x_2$ into SM particles. Thus the light component  forces the heavy one to keep its
equilibrium value, resulting in a  significant decrease of the relic density of $x_2$.
When both DM particles have similar masses, the interplay between the two sectors is more complicated, in particular the
 r\^{o}le of the interactions of the type $20\rightarrow 11$  will depend on the exact mass relation between the two DM particles.
For example,  this interaction can lead to an increase of the abundance of $x_2$ if $Y_1$ is large enough for the reverse process to give the largest contribution. 

\subsection{Numerical results}
The scalar model with a $Z_4$ symmetry contains two dark sectors. In sector 1 the DM candidate is a complex singlet, S, the main 
contribution to $\sigma_v^{1100}$ comes from annihilation  into Higgs pairs  and is determined by the term $\lambda_{S1} |S|^2 |H_1|^2$.
Sector 2 is similar to the Inert Doublet Model (IDM). The DM candidate can be either the scalar $H^0$ or the pseudoscalar $A^0$. Annihilation of  DM  
into SM particles is usually dominated by gauge boson pair production processes, while annihilation into fermion 
pairs as well as  co-annihilation processes can  also contribute. 
Furthermore, for  a DM mass at the electroweak scale, it was shown in~\cite{Yaguna:2010hn} that annihilation into 3-body final states via a virtual 
$W$ can be  important below the $W$ threshold. To avoid this complication we
will consider a DM with a mass above masses of the $W$, $Z$, and $h$. 
Under this condition, the DM annihilation   into SM particles in sector 2  is driven by $SU(2) \times U(1)$  gauge interactions and leads typically 
to a value of $\Omega h^2 < 0.1$, except for a  DM  heavier than about 500 GeV.
The co-annihilation of $H^0$, $A^0$, $H^+$ states increases $\Omega h^2$.

We will consider a benchmark point where both DM candidates $S$ and $H^0$ have a mass near 350 GeV. Other parameters are chosen so that semi-annihilation processes play an important role, while both components have comparable relic density and  $\Omega h^2=\Omega_1h^2+\Omega_2h^2=0.1$.   In particular to  have $\Omega_2 h^2 \approx 0.05 $ requires the contribution of coannihilation processes -- we therefore 
impose a small mass splitting $M_{H^0}\approx M_{A^0}$, meaning that $\lambda_5$ will be small, see Eq.~(\ref{eq:lam5:z4}).
Furthermore, a small value of $\lambda_4$ also leads to a small mass splitting with the charged Higgs. Note that for small $\lambda_5$ and $\lambda_4$  the positivity condition on the  potential, Eqs.~(\ref{eq:V:c},\ref{eq:Lag:3:Z:4}) is easily satisfied.

\begin{table}[hbt]
\begin{center}
\begin{tabular}{cc|cc|cc|cc}
$\lambda_2$     &    0.1   & $\lambda_{S1} $  &    0.1  & $\lambda'_{S}$&    0.1   & $M_{A}$  &  341~GeV \\
$\lambda_3$     &    0.1   & $\lambda_{S2} $  &    0.3  & $\mu_S$          &  100~GeV & $M_{H}$  &  339~GeV \\ 
$\lambda_4$     &    0.01  & $\lambda_{S12}$  &    0.13 & $ M_h$           &  125~GeV & $M_{S}$  &  350~GeV \\
$\lambda_S$     &    0.1   & $\lambda_{S21}$  &    0.13 &                  &          &          &          \\
\end{tabular}
\end{center}
\caption{Benchmark point for $Z_4$.}
\label{tab:bench:z4}
\end{table}

The results of the  calculation of the relic density when including different terms in 
Eq.~(\ref{z4eq1},\ref{z4eq2}) is presented in Table~\ref{tab:relic:z4}. 
When only (co-)annihilation into SM particles are taken into account, the relic density of $S$ is too high, 
while annihilation is much more efficient in Sector 2. Adding the interactions  of the type
of $1,1 \leftrightarrow 2,2$   brings the value of $\Omega_1 h^2$ and $\Omega_2 h^2$ closer to each other.
In our example the  DM in sector 1  has weak  interactions with
SM particles, therefore $\Omega_1 h^2$ is large when sector 2 is neglected.  As a  result of   interactions  with
sector 2 particles the  value for $\Omega_1 h^2$ is significantly reduced. This
effect was also observed for a DM model with  a $Z_2 \times Z_2$ symmetry
\cite{Belanger:2011ww}  and was called the {\it assisted freeze-out mechanism}. 
Finally, when semi-annihilation processes are included, both $\Omega_1 h^2$ and $\Omega_2 h^2$ decrease. 
\begin{table}[hbt]
\begin{center}
\begin{tabular}{lcc}
included terms of Eq(\ref{z4eq1},\ref{z4eq2})  &  $\Omega_1 h^2$  &  $\Omega_2 h^2$ \\
\hline
 $\sigma_v^{1100}$, $\sigma_v^{2200}$                           &    0.24 & 0.041  \\ 
 $\sigma_v^{1100}$, $\sigma_v^{2200}$ and $\sigma_v^{1122}$     &    0.079 & 0.064  \\ 
All                                                              & 0.050    & 0.051  \\ 
\end{tabular}
\end{center}
\caption{Relic density of DM particles for the $Z_4$ benchmark point.}
\label{tab:relic:z4}
\end{table}
Note that for this benchmark point, 
the cross section for DM elastic scattering on proton and neutron is $1.5(1.8) \cdot 10^{-9}$~pb  for the DM in sector 1 and 2, respectively. 
This is well below current exclusion limits of XENON100~\cite{Aprile:2011hi}, as will be discussed at the end of this section.

To examine more closely the interplay between the two DM sectors as well as  the role of semi-annihilation in determining the DM abundance, we let $M_S$ vary in the range 200-600~GeV and solve for the relic density by including new terms one by one. 
All other parameters are fixed to the value for the benchmark in Table~\ref{tab:bench:z4}. 

\begin{figure}
\begin{center}
 \includegraphics[height=8cm, width=15cm,angle=0]{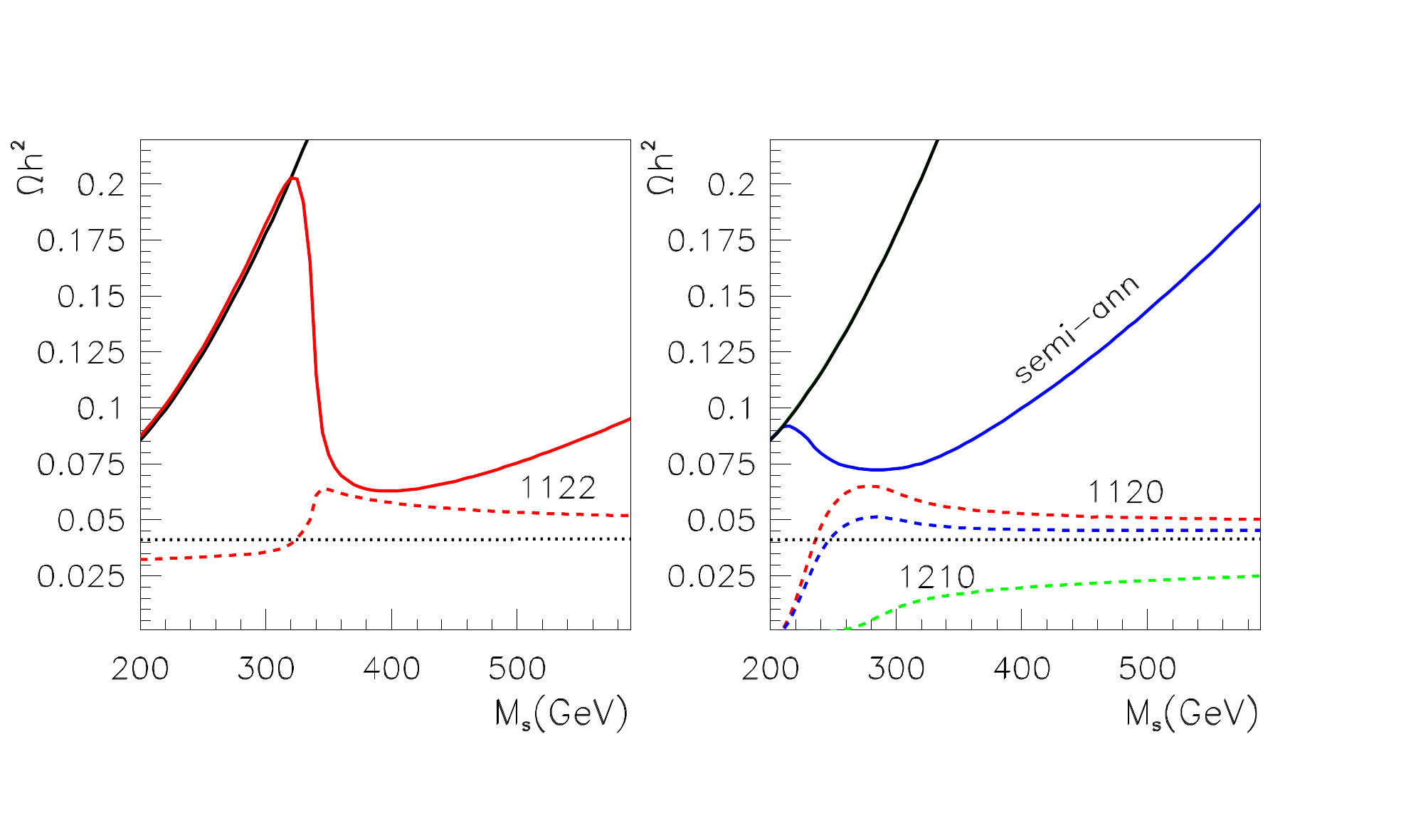}
 \vspace{-.8cm}
 \caption{ Effect of interactions between the two dark matter sectors (left) and of semi-annihilation (right) on $\Omega_1 h^2$(solid)  and $\Omega_2 h^2$(dashed)   as a function of $M_S$. 
 Left panel -- Including only $\sigma_v^{1100}$ and $\sigma_v^{2200}$(black) as well as $\sigma_v^{1122}$, $\sigma_v^{2211}$ (red).
 Right panel -- Including only $\sigma_v^{1210}$ (green), only $\sigma_v^{1120}$ (red) as well as all semi-annihilations  (blue), as a reference in black $\Omega_1 h^2$(solid) and $\Omega_2 h^2$ (dot) with only standard annihilation terms. Note that 
$\sigma_v^{1210}$ does not change $\Omega_1 h^2$.  }
\label{fig:omegaZ4:3}
\end{center}
\end{figure}

First we consider only the impact of annihilation processes, the results are displayed in Fig.~\ref{fig:omegaZ4:3}~(left).
When solving the evolution equation for the two DM independently, $\Omega_1 h^2$ rises rapidly with $M_S$ while   $\Omega_2 h^2 $ remains  constant. Note that for the model under consideration  we have that $\sigma_v^{1100}< \sigma_v^{2200}$ hence $\Omega_1 h^2>\Omega_2 h^2$.
The impact of $\sigma_v^{1122}$ and $\sigma_v^{2211}$ on the relic density depends on the relative masses of the scalar and doublet DM. 
The heavier DM candidate freezes out at a larger temperature than the lighter one, $T_{{\rm fo}\mathpzc{h}}>T_{{\rm fo}\mathpzc{l}}$.  
If the mass difference is large, this happens when the light one is at its equilibrium value. 
Thus  the contribution of $\sigma_v^{\mathpzc{h}\mathpzc{h}\mathpzc{l}\mathpzc{l}}$ just  adds to   $\sigma_v^{\mathpzc{h}\mathpzc{h}00}$, leading to a decrease of the  heavy DM  abundance.  
Furthermore, after  the light DM freezes out,
 interactions such as $\mathpzc{h}\mathpzc{h}\to \mathpzc{l}\mathpzc{l}$  give an additional source of light  DM, while the
reverse reaction is suppressed by a Boltzmann factor. This effect can, however, be small when the heavy particles have a  low density at this point.
 Thus  the interactions between DM sectors 1 and 2
 lead altogether to a decrease of the abundance of the heavy component and an increase  of the light component. 
This is observed in the left panel of Fig.~\ref{fig:omegaZ4:3}. In the region where $M_S<M_{H^0}=350$~GeV, $\Omega_2h^2$ decreases while in 
the region $M_S>M_{H^0}$, $\Omega_2h^2$ increases and vice-versa for $\Omega_1h^2$. 
Note that for large values of $M_S$, interactions with SM particles are weak so  $\sigma_v^{1122}\gg  \sigma_v^{1100}$,  leading to a large decrease in $\Omega_1h^2$. 
When there is a small difference between the two DM particles, the freeze-out temperatures of both component are similar. 
The density of the heavy DM component has not yet decreased to its final value at
the time the light component freezes out,  thus the effect of  $\mathpzc{h}\mathpzc{h} \leftrightarrow \mathpzc{l}\mathpzc{l}$ interactions
in increasing the abundance of the light component is more important. 
 This is particularly noticeable when looking at the curve for $\Omega_2 h^2$
 in the region, where $M_S$ is just above $M_{H^0}=350~{\rm GeV}$ in Fig.~\ref{fig:omegaZ4:3}~(left).
This discussion, where we ignore the semi-annihilation terms,
applies to models with $\lambda_{S12}=\lambda_{S21}=0$. In this case the $Z_4$ symmetry is replaced with a $Z_2 \times Z_2$ symmetry.

Next we consider the impact of semi-annihilation processes, ignoring the annihilation of pairs of particles from sector 1 to 2.  
The $\sigma_v^{1210}$ term  does not affect $\Omega_1 h^2$ and  
works as a catalyst for $ 2 \to \mathrm{SM}$ transitions. This term has an effect only after the freeze-out of $H^0$ and its effect is stronger when $Y_1$ is large, see Eq.~(\ref{z4eq2}).  Thus in the region  $M_S > M_{H^0}$ where the freeze-out of  $S$ occurs first (at a higher temperature), we
find a roughly  constant factor of suppression of $\Omega_2 h^2$. As $M_S$ decreases, its abundance  $Y_1$ at the freeze-out of $H^0$ ($T_{{\rm fo}\mathpzc{h}}$) will increase, thus the  suppression of  $\Omega_2 h^2$ is more important, see Fig.~\ref{fig:omegaZ4:3}.
Note that the suppression of $\Omega_2 h^2$ for  $M_S>  M_{H^0}$ is significantly larger than for the other semi-annihilation processes that we will discuss below.
This is because the $\sigma_v^{1210}$ term in Eq.~\ref{z4eq2} depends on $Y_1^2$, which is large  in this approximation.

\begin{figure}
\begin{center}
 \includegraphics[height=8cm, width=9cm,angle=0]{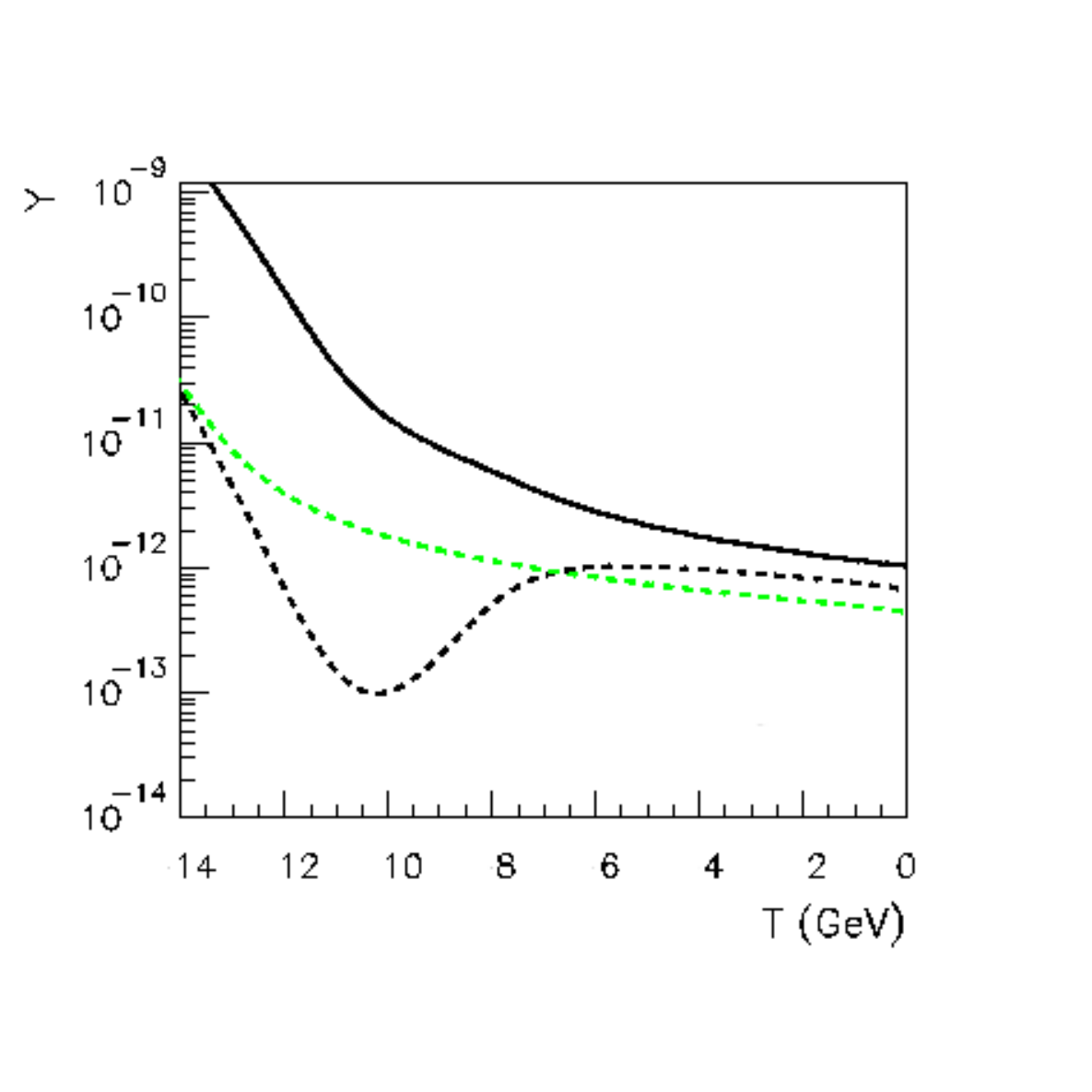}
 \vspace{-.8cm}
 \caption{   Temperature evolution of $Y_1$ (solid) and $Y_2$ (dashed) with standard terms and the contribution of  $\sigma_v^{1120}$ 
for $M_S=260$~GeV.   Temperature evolution of $Y_2$ with only standard terms (green/dashed), $T$ is in GeV.}
\label{fig:omegaZ4:4}
\end{center}
\end{figure}

The second type of  semi-annihilation process, $11\rightarrow 20$ (or its reverse $20\rightarrow 11$) leads to variations in the relic density of both DM components. 
If $M_S > M_{H^0}$, the impact of  $\sigma_v^{1120}$ is very similar to the one discussed above for  $\sigma_v^{1122}$. For $S$, the heavy component, 
the overall annihilation cross section is increased, leading to a  decrease in  $\Omega_1$, illustrated by the blue curve in Fig.~\ref{fig:omegaZ4:3}.
For $H^0$, the relic density  increases because the process $11 \to 20$ is an additional source of  sector 2 particles.
This increase is even more important when both particles have similar masses -- see the blue dashed curve in Fig.~\ref{fig:omegaZ4:3} when $M_S=260$-$350$~GeV.  To examine more closely the impact of the semi-annihilation in the region where the mass of both DM particles are similar, we compute the temperature evolution of $Y_1$ and $Y_2$ choosing
$M_S=260$~GeV. 
The result is displayed in Fig.~\ref{fig:omegaZ4:4}, in particular comparing  the evolution of $Y_2$ with and without 
the contribution of $\sigma_v^{1120}$. 
For this choice of masses,   the freeze-out of $H^0$ occurs when the 
abundance $Y_1=\overline{Y}_1$ is large, this means that  the term  
\begin{equation}
-\frac{1}{2}\sigma_v^{1120} \left( Y_1^2-\frac{Y_2}{\overline{Y}_2}{\overline{Y}_1^2} \right)  =
+\frac{1}{2}\sigma_v^{1120}\overline{Y}_1^2 \left( \frac{Y_2}{\overline{Y}_2} -1 \right) \ldots
\end{equation}
in Eq.~(\ref{z4eq2}) forces $Y_2$ to  follow its equilibrium value. Thus $Y_2$ is further reduced by semi-annihilation at large temperatures. 
After the freeze-out of  $S$, when $Y_1\gg \overline{Y}_1$, the same interaction leads to an increase of $Y_2$. 
Thus the overall effect is an increase in the abundance 
of class 2 particles as compared with the case where only standard interactions are considered.  

Finally, when $M_S<260~{\rm GeV}$,  the cross section  $\sigma_v^{1120}$, which consists of processes of the 
type $SS\rightarrow H^0 h$ is small  because of a lack of phase space, thus $\Omega_1 h^2$ is the same as when only standard annihilation terms were included. At the same time the reverse process, $20\rightarrow 11$ drives the depletion of class 2 particles and $\Omega_2 h^2$ drops to very small values. 
Note that when  $M_{H^0} > 2 M_S$  we expect that the class 2 DM will decay into pairs of class 1 particles since they are allowed 
by the  $Z_4$ symmetry.  However, in this example,  the effect of 
$\sigma_v^{1210}$ and $\sigma_v^{1120}$ terms already leads to  very small values of $\Omega_2 h^2$ for low values of $M_S$, so that the decays are irrelevant. In summary, 
the combined effect of semi-annihilation processes is for this example close to the result of only including $\sigma_v^{1120}$, see Fig.~\ref{fig:omegaZ4:3}. 

\begin{figure}
\begin{center}
 \includegraphics[height=8cm, width=7.8cm,angle=0]{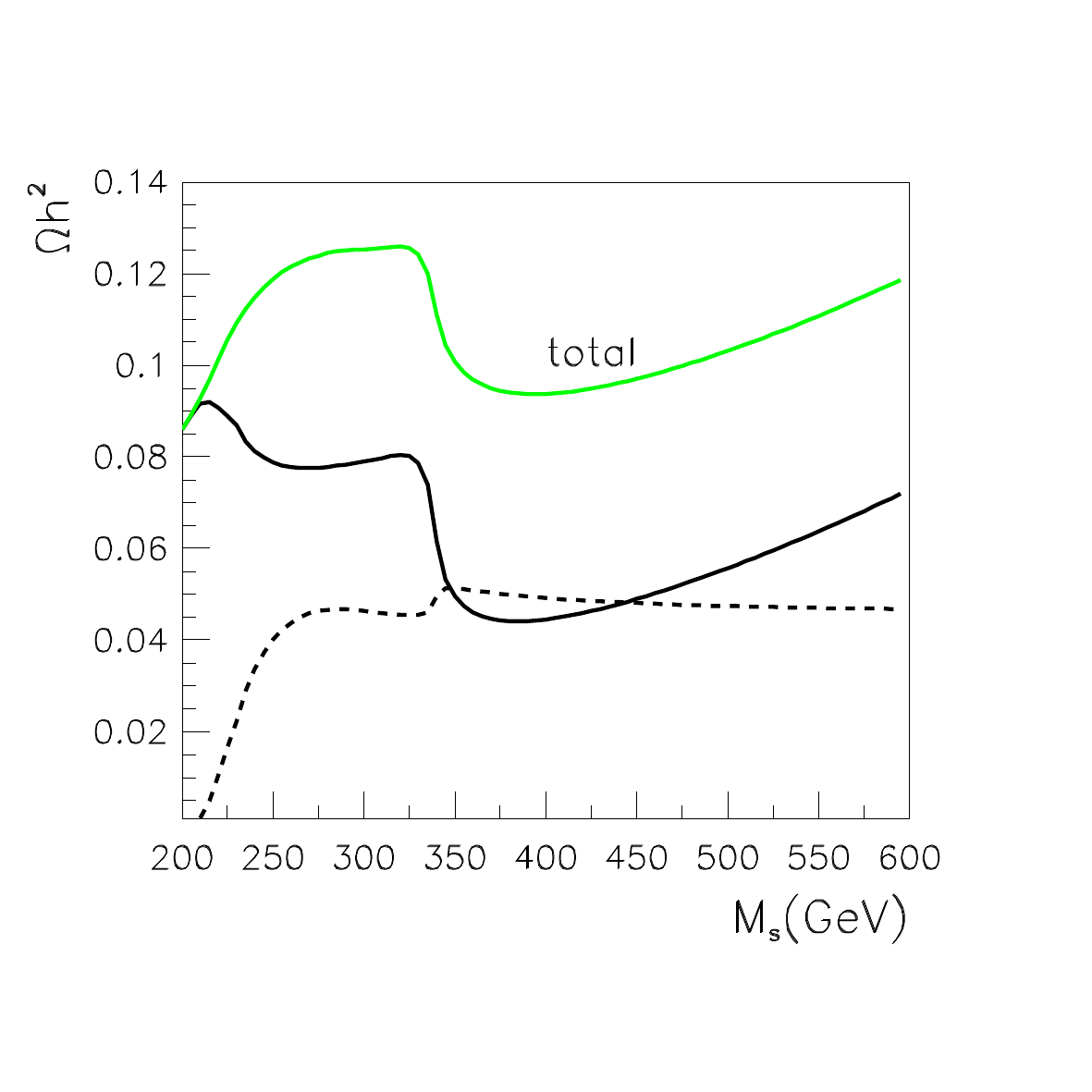}
  \includegraphics[height=8cm, width=7.8cm,angle=0]{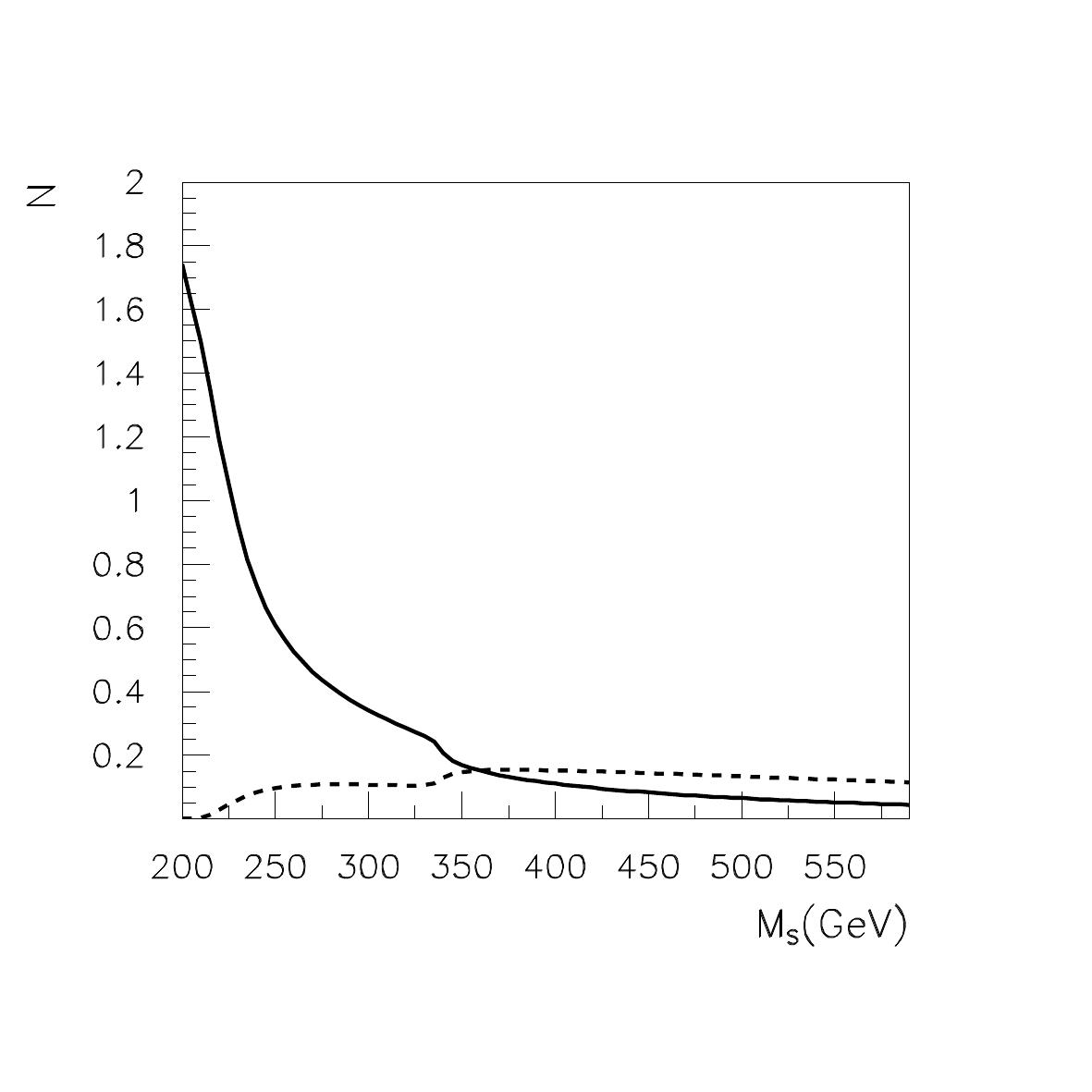}
 \vspace{-.8cm}
 \caption{ Left: $\Omega_1 h^2$(solid), $\Omega_2 h^2$(dashed)   and $\Omega h^2$ (green) as a function of $M_S$, the singlet DM mass. Right: Number of events expected in XENON100 from $S$ (solid) and $H^0$ (dashed) elastic scattering as a function of $M_S$.
}
\label{fig:omegaZ4}
\end{center}
\end{figure}

The result for $\Omega_1h^2$ and $\Omega_2h^2$ including all annihilation and semi-annihilation processes 
is displayed in Fig.~\ref{fig:omegaZ4}. The semi-annihilation mechanisms dominate for $M_S<M_{H^0}$ while the assisted freeze-out mechanism is the dominant effect when $M_S>M_{H^0}$. The total dark matter abundance is within about 10\% of the value preferred by WMAP measurements over the whole range of masses considered. 
While the features we have described here are generic, the relative importance of different annihilation and semi-annihilation processes is model dependent and depends on the size of the various cross sections within a specific model.

Finally we compute the spin-independent cross section for $S$ and $H^0$ scattering on xenon nuclei. 
As mentioned above at the benchmark point $\sigma^{\rm SI}=1.5(1.8) \times10^{-9}$~pb  for the DM in sector 1 and 2 respectively. We then compute the number of events that should be expected in XENON100 ~\cite{Aprile:2011hi} in the interval $8.4~{\rm keV}< E< 44.6~{\rm keV}$ after an exposure of $1171~{\rm kg} \cdot{\rm day}$. 
The number of events is directly proportional to the DM local density and we assume that the fraction of each DM component locally is the same as in the early universe, $\rho_i=\rho \Omega_i/\Omega_{\rm tot}$ where $\rho=0.3$.
For $S$ the cross section is largest for small masses, furthermore $S$ contributes maximally to the DM density, hence the maximum predicted number of events, see Fig.~\ref{fig:omegaZ4}. The cross section for $H^0$ scattering on nuclei is clearly independent of $M_S$, the variation of the number of events is simply due to the variation in the density of the second DM.

\section{CONCLUSIONS}
We have formulated scalar dark matter models with the minimal particle content in which dark matter stability is due to the discrete $Z_N$ symmetry with $N>2.$
Already the minimal models containing one extra scalar singlet and doublet possess non-trivial dark matter phenomenology.
In particular, the annihilation processes with new topologies like $x_ix_j \rightarrow x_k X$, where $x_i$ is one of the dark matter particles and $X$ is any standard 
model particle,
change the dark matter freeze-out process and must be taken into account when calculating the dark matter relic abundance. Furthermore, in models with two dark matter candidates, annihilation processes involving only particles of two different dark matter sectors also impact the relic abundance of both dark matter particles.
We have performed an example study of  semi-annihilations in two scalar dark matter models based on $Z_3$ and $Z_4$ symmetries. 
We  implemented those models for micrOMEGAs and 
studied  the impact of semi-annihilations and of the interactions between the dark sectors on the generation of dark matter relic abundance at the early Universe and  the predictions for dark matter direct detection relevant for the presently running XENON100 experiment.
We conclude that in this type of models both semi-annihilations and dark sector interactions may significantly affect the dark matter phenomenology compared to the 
well studied $Z_2$ models, and, therefore,  must be taken into account  in precise numerical analyses of dark matter properties.

\section*{ACKNOWLEDGEMENTS}
Part of this work was performed in the  Les Houches 2011 {\it Physics at TeV colliders} Workshop. K.K. and M.R. were supported by the ESF grants 8090, 8499, 8943, MTT8, MTT59, MTT60, MJD140, by the recurrent financing SF0690030s09 project
and by the European Union through the European Regional Development Fund.
A.P. was supported by the Russian foundation for Basic Research, grant RFBR-10-02-01443-a. The work of A.P. and G.B. was supported in part by the GDRI-ACPP of CNRS.

\bibliography{Z34DM}

\providecommand{\href}[2]{#2}\begingroup\raggedright\begin{thebibliography}{10}

\bibitem{Farrar:1978xj}
G.~R. Farrar and P.~Fayet, {\it {Phenomenology of the Production, Decay, and
  Detection of New Hadronic States Associated with Supersymmetry}},  {\em
  Phys.Lett.} {\bf B76} (1978) 575--579.

\bibitem{Krauss:1988zc}
L.~M. Krauss and F.~Wilczek, {\it {Discrete Gauge Symmetry in Continuum
  Theories}},  {\em Phys.Rev.Lett.} {\bf 62} (1989) 1221.

\bibitem{Fritzsch:1974nn}
H.~Fritzsch and P.~Minkowski, {\it {Unified Interactions of Leptons and
  Hadrons}},  {\em Annals Phys.} {\bf 93} (1975) 193--266.

\bibitem{Martin:1992mq}
S.~P. Martin, {\it {Some simple criteria for gauged R-parity}},  {\em
  Phys.Rev.} {\bf D46} (1992) 2769--2772,
  [\href{http://xxx.lanl.gov/abs/hep-ph/9207218}{{\tt hep-ph/9207218}}].

\bibitem{Kadastik:2009dj}
M.~Kadastik, K.~Kannike, and M.~Raidal, {\it {Matter parity as the origin of
  scalar Dark Matter}},  {\em Phys.Rev.} {\bf D81} (2010) 015002,
  [\href{http://xxx.lanl.gov/abs/0903.2475}{{\tt arXiv:0903.2475}}].

\bibitem{Kadastik:2009cu}
M.~Kadastik, K.~Kannike, and M.~Raidal, {\it {Dark Matter as the signal of
  Grand Unification}},  {\em Phys.Rev.} {\bf D80} (2009) 085020,
  [\href{http://xxx.lanl.gov/abs/0907.1894}{{\tt arXiv:0907.1894}}].

\bibitem{Ibanez:1991hv}
L.~E. Ibanez and G.~G. Ross, {\it {Discrete gauge symmetry anomalies}},  {\em
  Phys.Lett.} {\bf B260} (1991) 291--295.

\bibitem{Agashe:2004ci}
K.~Agashe and G.~Servant, {\it {Warped unification, proton stability and dark
  matter}},  {\em Phys.Rev.Lett.} {\bf 93} (2004) 231805,
  [\href{http://xxx.lanl.gov/abs/hep-ph/0403143}{{\tt hep-ph/0403143}}].

\bibitem{Agashe:2004bm}
K.~Agashe and G.~Servant, {\it {Baryon number in warped GUTs: Model building
  and (dark matter related) phenomenology}},  {\em JCAP} {\bf 0502} (2005) 002,
  [\href{http://xxx.lanl.gov/abs/hep-ph/0411254}{{\tt hep-ph/0411254}}].

\bibitem{Dreiner:2005rd}
H.~K. Dreiner, C.~Luhn, and M.~Thormeier, {\it {What is the discrete gauge
  symmetry of the MSSM?}},  {\em Phys.Rev.} {\bf D73} (2006) 075007,
  [\href{http://xxx.lanl.gov/abs/hep-ph/0512163}{{\tt hep-ph/0512163}}].

\bibitem{Ma:2007gq}
E.~Ma, {\it {Z(3) Dark Matter and Two-Loop Neutrino Mass}},  {\em Phys.Lett.}
  {\bf B662} (2008) 49--52, [\href{http://xxx.lanl.gov/abs/0708.3371}{{\tt
  arXiv:0708.3371}}].

\bibitem{Agashe:2010gt}
K.~Agashe, D.~Kim, M.~Toharia, and D.~G. Walker, {\it {Distinguishing Dark
  Matter Stabilization Symmetries Using Multiple Kinematic Edges and Cusps}},
  {\em Phys.Rev.} {\bf D82} (2010) 015007,
  [\href{http://xxx.lanl.gov/abs/1003.0899}{{\tt arXiv:1003.0899}}].

\bibitem{Agashe:2010tu}
K.~Agashe, D.~Kim, D.~G. Walker, and L.~Zhu, {\it {Using $M_{T2}$ to
  Distinguish Dark Matter Stabilization Symmetries}},  {\em Phys.Rev.} {\bf
  D84} (2011) 055020, [\href{http://xxx.lanl.gov/abs/1012.4460}{{\tt
  arXiv:1012.4460}}].

\bibitem{Batell:2010bp}
B.~Batell, {\it {Dark Discrete Gauge Symmetries}},  {\em Phys.Rev.} {\bf D83}
  (2011) 035006, [\href{http://xxx.lanl.gov/abs/1007.0045}{{\tt
  arXiv:1007.0045}}].

\bibitem{DEramo:2010ep}
F.~D'Eramo and J.~Thaler, {\it {Semi-annihilation of Dark Matter}},  {\em JHEP}
  {\bf 1006} (2010) 109, [\href{http://xxx.lanl.gov/abs/1003.5912}{{\tt
  arXiv:1003.5912}}].

\bibitem{Liu:2011aa}
Z.-P. Liu, Y.-L. Wu, and Y.-F. Zhou, {\it {Enhancement of dark matter relic
  density from the late time dark matter conversions}},  {\em Eur.Phys.J.} {\bf
  C71} (2011) 1749, [\href{http://xxx.lanl.gov/abs/1101.4148}{{\tt
  arXiv:1101.4148}}].

\bibitem{Belanger:2011ww}
G.~Belanger and J.-C. Park, {\it {Assisted freeze-out}},
  \href{http://xxx.lanl.gov/abs/1112.4491}{{\tt arXiv:1112.4491}}.

\bibitem{Belanger:2004yn}
G.~Belanger, F.~Boudjema, A.~Pukhov, and A.~Semenov, {\it {micrOMEGAs: Version
  1.3}},  {\em Comput.Phys.Commun.} {\bf 174} (2006) 577--604,
  [\href{http://xxx.lanl.gov/abs/hep-ph/0405253}{{\tt hep-ph/0405253}}].

\bibitem{Belanger:2006is}
G.~Belanger, F.~Boudjema, A.~Pukhov, and A.~Semenov, {\it {MicrOMEGAs 2.0: A
  Program to calculate the relic density of dark matter in a generic model}},
  {\em Comput.Phys.Commun.} {\bf 176} (2007) 367--382,
  [\href{http://xxx.lanl.gov/abs/hep-ph/0607059}{{\tt hep-ph/0607059}}].

\bibitem{Deshpande:1977rw}
N.~G. Deshpande and E.~Ma, {\it {Pattern of Symmetry Breaking with Two Higgs
  Doublets}},  {\em Phys.Rev.} {\bf D18} (1978) 2574.

\bibitem{Ma:2006km}
E.~Ma, {\it {Verifiable radiative seesaw mechanism of neutrino mass and dark
  matter}},  {\em Phys.Rev.} {\bf D73} (2006) 077301,
  [\href{http://xxx.lanl.gov/abs/hep-ph/0601225}{{\tt hep-ph/0601225}}].

\bibitem{Barbieri:2006dq}
R.~Barbieri, L.~J. Hall, and V.~S. Rychkov, {\it {Improved naturalness with a
  heavy Higgs: An Alternative road to LHC physics}},  {\em Phys.Rev.} {\bf D74}
  (2006) 015007, [\href{http://xxx.lanl.gov/abs/hep-ph/0603188}{{\tt
  hep-ph/0603188}}].

\bibitem{LopezHonorez:2006gr}
L.~Lopez~Honorez, E.~Nezri, J.~F. Oliver, and M.~H. Tytgat, {\it {The Inert
  Doublet Model: An Archetype for Dark Matter}},  {\em JCAP} {\bf 0702} (2007)
  028, [\href{http://xxx.lanl.gov/abs/hep-ph/0612275}{{\tt hep-ph/0612275}}].

\bibitem{McDonald:1993ex}
J.~McDonald, {\it {Gauge Singlet Scalars as Cold Dark Matter}},  {\em Phys.
  Rev.} {\bf D50} (1994) 3637--3649,
  [\href{http://xxx.lanl.gov/abs/hep-ph/0702143}{{\tt hep-ph/0702143}}].

\bibitem{Barger:2007im}
V.~Barger, P.~Langacker, M.~McCaskey, M.~J. Ramsey-Musolf, and G.~Shaughnessy,
  {\it {LHC Phenomenology of an Extended Standard Model with a Real Scalar
  Singlet}},  {\em Phys. Rev.} {\bf D77} (2008) 035005,
  [\href{http://xxx.lanl.gov/abs/0706.4311}{{\tt arXiv:0706.4311}}].

\bibitem{Barger:2008jx}
V.~Barger, P.~Langacker, M.~McCaskey, M.~Ramsey-Musolf, and G.~Shaughnessy,
  {\it {Complex Singlet Extension of the Standard Model}},  {\em Phys. Rev.}
  {\bf D79} (2009) 015018, [\href{http://xxx.lanl.gov/abs/0811.0393}{{\tt
  arXiv:0811.0393}}].

\bibitem{Burgess:2000yq}
C.~P. Burgess, M.~Pospelov, and T.~ter Veldhuis, {\it {The minimal model of
  nonbaryonic dark matter: A singlet scalar}},  {\em Nucl. Phys.} {\bf B619}
  (2001) 709--728, [\href{http://xxx.lanl.gov/abs/hep-ph/0011335}{{\tt
  hep-ph/0011335}}].

\bibitem{Gonderinger:2009jp}
M.~Gonderinger, Y.~Li, H.~Patel, and M.~J. Ramsey-Musolf, {\it {Vacuum
  Stability, Perturbativity, and Scalar Singlet Dark Matter}},  {\em JHEP} {\bf
  1001} (2010) 053, [\href{http://xxx.lanl.gov/abs/0910.3167}{{\tt
  arXiv:0910.3167}}].

\bibitem{Kadastik:2009ca}
M.~Kadastik, K.~Kannike, A.~Racioppi, and M.~Raidal, {\it {EWSB from the soft
  portal into Dark Matter and prediction for direct detection}},  {\em
  Phys.Rev.Lett.} {\bf 104} (2010) 201301,
  [\href{http://xxx.lanl.gov/abs/0912.2729}{{\tt arXiv:0912.2729}}].

\bibitem{Kadastik:2009gx}
M.~Kadastik, K.~Kannike, A.~Racioppi, and M.~Raidal, {\it {Implications of the
  CDMS result on Dark Matter and LHC physics}},  {\em Phys.Lett.} {\bf B694}
  (2010) 242--245, [\href{http://xxx.lanl.gov/abs/0912.3797}{{\tt
  arXiv:0912.3797}}].

\bibitem{Huitu:2010uc}
K.~Huitu, K.~Kannike, A.~Racioppi, and M.~Raidal, {\it {Long-lived charged
  Higgs at LHC as a probe of scalar Dark Matter}},  {\em JHEP} {\bf 1101}
  (2011) 010, [\href{http://xxx.lanl.gov/abs/1005.4409}{{\tt
  arXiv:1005.4409}}].

\bibitem{Gondolo:1990dk}
P.~Gondolo and G.~Gelmini, {\it {Cosmic abundances of stable particles:
  Improved analysis}},  {\em Nucl.Phys.} {\bf B360} (1991) 145--179.

\bibitem{Belanger:2010gh}
G.~Belanger, F.~Boudjema, P.~Brun, A.~Pukhov, S.~Rosier-Lees, {\em et.~al.},
  {\it {Indirect search for dark matter with micrOMEGAs2.4}},  {\em
  Comput.Phys.Commun.} {\bf 182} (2011) 842--856,
  [\href{http://xxx.lanl.gov/abs/1004.1092}{{\tt arXiv:1004.1092}}].

\bibitem{Aprile:2011hi}
{\bf XENON100 Collaboration} Collaboration, E.~Aprile {\em et.~al.}, {\it {Dark
  Matter Results from 100 Live Days of XENON100 Data}},  {\em Phys.Rev.Lett.}
  {\bf 107} (2011) 131302, [\href{http://xxx.lanl.gov/abs/1104.2549}{{\tt
  arXiv:1104.2549}}].

\bibitem{Belanger:2010cd}
G.~Belanger, M.~Kakizaki, E.~Park, S.~Kraml, and A.~Pukhov, {\it {Light mixed
  sneutrinos as thermal dark matter}},  {\em JCAP} {\bf 1011} (2010) 017,
  [\href{http://xxx.lanl.gov/abs/1008.0580}{{\tt arXiv:1008.0580}}].

\bibitem{Belanger:2001fz}
G.~Belanger, F.~Boudjema, A.~Pukhov, and A.~Semenov, {\it {MicrOMEGAs: A
  Program for calculating the relic density in the MSSM}},  {\em
  Comput.Phys.Commun.} {\bf 149} (2002) 103--120,
  [\href{http://xxx.lanl.gov/abs/hep-ph/0112278}{{\tt hep-ph/0112278}}].

\bibitem{Yaguna:2010hn}
C.~E. Yaguna, {\it {Large contributions to dark matter annihilation from
  three-body final states}},  {\em Phys.Rev.} {\bf D81} (2010) 075024,
  [\href{http://xxx.lanl.gov/abs/1003.2730}{{\tt arXiv:1003.2730}}].

\end{thebibliography}\endgroup

\end{document}